\documentclass[10pt,reqno]{amsart}
\usepackage{amssymb}
\usepackage{amsfonts}
\usepackage{amsthm}
\usepackage{amsmath}
\usepackage{epsfig}

\theoremstyle{definition}

\newcommand{\IR}{{\mathbb{R}}}

\newcommand{\ZZ}{{\mathbb{Z}}}

\newcommand{\IQ}{{\mathbb{Q}}}
\newcommand{\Exp}{{\mathbb{E}\text{xp}}}
\let\det=\undefined\DeclareMathOperator*{\det}{det}
\let\Re=\undefined\DeclareMathOperator*{\Re}{Re}

\DeclareMathOperator*{\tr}{tr} \DeclareMathOperator*{\diag}{diag}

\title[Universality for mathematical and physical systems]{Universality for mathematical and\\ physical systems}

\author[Percy Deift]{Percy Deift}
\address{Percy Deift\\Courant Institute of Mathematical Sciences\\ New York University\\
251 Mercer Street\\ New York, NY 10012}
\email{deift@cims.nyu.edu}

\thanks{The author would like to thank Sourav Chatterjee, Patrik Ferrari, and Peter Sarnak for useful
comments and Irina Nenciu for her help and suggestions in preparing the manuscript. The work of the author
was supported in part by DMS Grants No. 0296084 and No. 0500923, and also by a Friends of the Institute Visiting Membership at the
Institute for Advanced Study in Princeton, Spring 2006.}

\begin{document}

\begin{abstract}
All physical systems in equilibrium obey the laws of thermodynamics. In other words, whatever the precise nature of the
interaction between the atoms and molecules at the microscopic level, at the macroscopic level, physical systems exhibit universal behavior in
the sense that they are all governed by the same laws and formulae of thermodynamics.
In this paper we describe some recent history of universality ideas in physics starting with Wigner's model for the scattering of neutrons
off large nuclei and show how these ideas have led mathematicians to investigate universal behavior for a variety of mathematical systems.
This is true not only for systems which have a physical origin, but also
for systems which arise in a purely mathematical context such as the Riemann hypothesis, and a version of the card game solitaire called patience
sorting.
\end{abstract}



\maketitle

\section{Introduction}

All physical systems in equilibrium obey the laws of thermodynamics.
The first law asserts the conservation of energy. The second law
has a variety of formulations, one of which is the
following: Suppose that in a work cycle a heat engine extracts $Q_1$
units of heat from a heat reservoir at temperature $T_1$, performs
$W$ units of work, and then exhausts the remaining $Q_2=Q_1-W$ units
of heat to a heat sink at temperature $T_2<T_1$. Let
$\eta=\frac{W}{Q_1}$ denote the efficiency of the conversion of heat
into work. Then the second law tells us there is a maximal
efficiency $\eta_\text{max}=(T_1-T_2)/T_1$, depending only on
$T_1$ and $T_2$, so that for all heat engines, and all work cycles,
\begin{equation}\label{E:1}
\eta\leq \eta_\text{max}\,.
\end{equation}
Nature is so set up that we just cannot do any better.

On the other hand, it is a very old thought, going back at least to
Democritus and the Greeks, that matter, all matter, is built out of
tiny constituents - atoms - obeying their own laws of interaction.
The juxtaposition of these two points of view, the macroscopic world
of tangible objects and the microscopic world of atoms, presents a
fundamental, difficult and long-standing challenge to scientists;
namely, how does one derive the macroscopic laws of thermodynamics
from the microscopic laws of atoms? The special, salient feature of
this challenge is that the \textit{same} laws of thermodynamics
should emerge no matter what the details of the atomic interaction.
In other words, on the macroscopic scale, physical systems should
exhibit universality\footnote{In physics, the term ``universality''
is usually used in the more limited context of scaling laws for
critical phenomena. In this paper we use the term
``universality'' more broadly in the spirit of the preceding
discussion. We trust this will cause no confusion.}. Indeed,
it is the very emergence of universal behavior for
macroscopic systems that makes possible the existence of physical
laws.

This kind of thinking, however, is not common in the world of
mathematics. Mathematicians tend to think of their problems as sui
generis, each with its own special, distinguishing features. Two
problems are regarded as ``the same'' only if some isomorphism,
explicit or otherwise, can be constructed between them. In recent
years, however, universality in the above sense of macroscopic
physics has started to emerge in a wide variety of mathematical
problems, and the goal of this paper is to illustrate some of these
developments. As we will see, there are problems from diverse areas, often
with no discernible, mechanistic connections, all of which behave,
on the appropriate scale, in precisely the same way. The list of
such problems is varied, long and growing, and points to the
emergence of what one might call ``macroscopic mathematics.''

A precedent for the kind of results that we are going to describe is
given by the celebrated central limit theorem of probability theory,
where one considers independent, identically distributed variables
$\{x_n\}_{n\geq1}$. The central limit theorem tells us that if we
center and scale the variables, $x_n\to y_n\equiv
\bigl(x_n-\Exp(x_n)\bigr)/\sqrt{\mathbb{V}\text{ar}(x_n)}$, then
\begin{equation}\label{E:2}
\lim_{n\to\infty} \text{Prob}\left(\frac{\sum_{k=1}^n y_k}{\sqrt
n}\leq t\right) =\int_{-\infty}^t
e^{-\tfrac{u^2}{2}}\,\frac{du}{\sqrt{2\pi}}\,.
\end{equation}
We see here explicitly that the Gaussian distribution on the
right-hand side of \eqref{E:2} is \textit{universal}, independent of
the distribution for the $x_n$'s. The proof of the central limit
theorem for independent coin flips,
$\text{Prob}(x_n=+1)=\text{Prob}(x_n=-1)=\frac12$, goes back to de
Moivre and Laplace in the $18^\text{th}$ century.

The outline of the paper is as follows: In Section \ref{S:2} we will
introduce and discuss some models from random matrix theory (RMT).
Various distributions associated with these models will play the
same role in the problems that we discuss later on in the paper as
the Gaussian does in \eqref{E:2}. As noted above, thermodynamics
reflects universality for all macroscopic systems, but there are
also many universality sub-classes which describe the behavior of
physical systems in restricted situations. For example, many fluids,
such as water and vinegar, obey the Navier-Stokes equation, but a
variety of heavy oils obey the lubrication equations. In the same
way we will see that certain mathematical problems are described by
so-called Unitary Ensembles of random matrices, and others by
so-called Orthogonal or Symplectic Ensembles. In Section \ref{S:3},
we present a variety of problems from different areas of
mathematics, and in Section \ref{S:4} we show how these problems are
described by random matrix models from Section \ref{S:2}. In the
final Section \ref{S:5} we discuss briefly some of the mathematical
methods that are used to prove the results in Section \ref{S:4}.
Here combinatorial identities, Riemann-Hilbert problems (RHP's) and the nonlinear steepest
descent method of \cite{DeiZho1}, as well as the classical steepest descent method, play a key role. We end the
section with some speculations, suggesting how to place the results
of Sections \ref{S:3} and \ref{S:4} in a broader mathematical
framework.

\section{Random matrix models}\label{S:2}

There are many ensembles of random matrices that are of interest,
and we refer the reader to the classic text of Mehta~\cite{Meh} for
more information (see also \cite{Dei1}). In this paper we will
consider almost exclusively (see, however, \eqref{E:48} et seq. below) only three
kinds of ensembles:
\begin{itemize}
  \item[(a)] Orthogonal Ensembles (OE's) consisting of $N\times N$ real symmetric
  matrices $M$, $M=\bar M=M^T$.
  \item[(b)] Unitary Ensembles (UE's) consisting of $N\times N$ Hermitian matrices $M$,
  $M=M^*$.
  \item[(c)] Symplectic Ensembles (SE's) consisting of $2N\times 2N$ Hermitian, self-dual matrices
  $M=M^*=JM^TJ^T$, where $J$ is the standard $2N\times 2N$ block diagonal symplectic matrix,
  $J=\diag (\tau,\tau,\dots,\tau)$, $\tau=\left(\begin{matrix}0&1\\-1&0\end{matrix}\right)$.
\end{itemize}
For reasons that will soon become clear, OE's, UE's and SE's are
labeled by a parameter $\beta$, where $\beta=1,2$ or 4,
respectively. In all three cases the ensembles are equipped with
probability distributions of the form
\begin{equation}\label{E:3}
P_{N,\beta}(M)\,d_\beta M=\frac{1}{Z_{N,\beta}}
e^{-\tr\bigl(V_{N,\beta}(M)\bigr)}\,d_\beta M
\end{equation}
where $V_{N,\beta}$ is a real-valued function on $\IR$ such that
$V_{N,\beta}(x)\to+\infty$ sufficiently rapidly as $|x|\to\infty$,
$Z_{N,\beta}$ is a normalization coefficient, and $d_\beta M $
denotes Lebesgue measure on the algebraically independent entries of
$M$. For example, in the orthogonal case, $d_{\beta=1}M=\prod_{1\leq
j\leq k\leq N} dM_{jk}$, where $M=(M_{jk})$ (see, e.g. \cite{Meh}).
The notation ``orthogonal'', ``unitary'', and ``symplectic'' refers
to the fact that the above ensembles with associated distributions
\eqref{E:3} are invariant under conjugation $M\to SMS^{-1}$, where
$S$ is orthogonal, unitary, or unitary-symplectic (i.e., $S\in
USp(2N)=\{S\,:\,SS^*=I,\,SJS^T=J\}$) respectively. When
$V_{N,\beta}(x)=x^2$, one has the so-called Gaussian Orthogonal
Ensemble (GOE), the Gaussian Unitary Ensemble (GUE), and the
Gaussian Symplectic Ensemble (GSE), for $\beta=1,2$ or 4,
respectively.

The distributions \eqref{E:3} give rise in turn to distributions on
the eigenvalues $\lambda_1\leq\lambda_2\leq\cdots$ of $M$
\begin{equation}\label{E:4}
\hat P_{N,\beta}(\lambda) \, d^N\lambda =\frac{1}{\hat Z_{N,\beta}}
e^{-\eta_\beta \sum_{i=1}^N V_{N,\beta}(\lambda_i)} \prod_{1\leq
i<j\leq N}|\lambda_i-\lambda_j|^\beta\,d\lambda_1\cdots d\lambda_N
\end{equation}
where $\hat Z_{N,\beta}$ is again a normalization coefficient, and
$\eta_\beta=1$ if $\beta=1$ or 2 and $\eta_4=2$ (this is because the
eigenvalues for $\beta=4$ double up). The labeling of OE's, UE's,
and SE's by $\beta=1,2$ and 4 is now clear. In all three cases, we
see that the random matrix ensembles give rise to random particle
systems $\{\lambda_1,\lambda_2,\ldots\}$ with \textit{repulsion}
built in: the probability that two eigenvalues are close together is
small and vanishes like a power of the distance between them. This
is an essential feature of random matrix ensembles, in contrast to
random Poisson particle systems, say, where the particles may bunch
together or exhibit large gaps.

Loosely speaking, we say that a system is \textit{modeled by random
matrix theory (RMT) if it behaves statistically like the eigenvalues
of a ``large'' OE, UE,... matrix}. In analyzing such systems there
is something known as the \textit{standard procedure}: Suppose we
wish to compare some statistical quantities $\{a_k\}$ in the
neighborhood of some point $A$ with the eigenvalues $\{\lambda_k\}$
of some matrix in the neighborhood of some energy $E$, say, in the
bulk of the spectrum. Then we always \textit{center} and
\textit{scale} the $a_k$'s and the $\lambda_k$'s,
\begin{equation}\label{E:5}
a_k\to\tilde a_k=\gamma_a(a_k-A),\qquad
\lambda_k\to\tilde\lambda_k=\gamma_\lambda(\lambda_k-E)
\end{equation}
so that
\begin{equation}\label{E:6}
\Exp\bigl(\#\{\tilde a_k\text{'s per unit interval}\}\bigr)
=\Exp\bigl(\#\{\tilde\lambda_k\text{'s per unit interval}\}\bigr)=1.
\end{equation}
For energies $E$ at the edge of the spectrum, the above procedure
must be modified slightly (see below).

This procedure can be viewed as follows: A scientist wishes to
investigate some statistical phenomenon. What s'he has at hand is a
microscope and a handbook of matrix ensembles. The data $\{a_k\}$
are embedded on a slide which can be inserted into the microscope.
The only freedom that the scientist has is to center the slide,
$a_k\to a_k-A$, and then adjust the focus $a_k-A\to\tilde
a_k=\gamma_a(a_k-A)$ so that on average one data point $\tilde a_k$
appears per unit length on the slide. At that point the scientist
takes out his'r handbook, and then tries to match the statistics of
the $\tilde a_k$'s with those of the eigenvalues of some ensemble.
If the fit is good, the scientist then says that the system is
well-modeled by RMT.

It is a remarkable fact, going back to the work of Gaudin and Mehta,
and later Dyson, in the 1960's, that the key statistics for OE's,
UE's, and SE's can be computed in closed form. This is true not only
for finite $N$, but also for various scaling limits as $N\to\infty$.
For GOE, GUE, and GSE we refer the reader to \cite{Meh}. Here the
Hermite polynomials, which are orthogonal with respect to the weight
$e^{-x^2}\,dx$ on $\IR$, play a critical role, and the scaling
limits as $N\to\infty$ follow from the known, classical asymptotics
of the Hermite polynomials. For UE's with general potentials
$V_{N,\beta=2}$, the techniques described in \cite{Meh} for GUE go
through for finite $N$, the role of the Hermite polynomials now
being played by the polynomials orthogonal with respect to the
weight $e^{-V_{N,\beta=2}(x)}\,dx$ on $\IR$ (see, e.g. \cite{Dei1}).
For general $V_{N,\beta=2}$, however, the asymptotic behavior of
these polynomials as $N\to\infty$ does not follow from classical
estimates. In order to overcome this obstacle, the authors in
\cite{DKMVZ1} and \cite{DKMVZ2} (see also \cite{Dei1} for a
pedagogical presentation) used the Riemann-Hilbert steepest-descent
method introduced by Deift and Zhou \cite{DeiZho1}, and further
developed with Venakides \cite{DVZ}, to compute the asymptotics as
$N\to\infty$ of the orthogonal polynomials for a very general class
of analytic weights. In view of the preceding comments, the scaling
limits of the key statistics for UE's then follow for such weights
(see also \cite{BI} for the special case
$V_{N,\beta=2}(x)=N(x^4-tx^2)$).
For another approach to UE universality, see \cite{PasSch}.
For OE's and SE's with classical
weights, such as Laguerre, Jacobi, etc., for which the asymptotics
of the associated orthogonal polynomials are known, the GOE and GSE
methods in \cite{Meh} apply (see the introductions to \cite{DG1} and
\cite{DG2} for a historical discussion). For general $V_{N,\beta}$,
$\beta=1$ or 4, new techniques are needed, and these were
introduced, for finite $N$, by Tracy and Widom in \cite{TW1} and
\cite{W}. In \cite{DG1} and \cite{DG2}, the authors use the results
in \cite{TW1} and \cite{W}, together with the asymptotic estimates
in \cite{DKMVZ2}, to compute the large $N$ limits of the key
statistics for OE's and SE's with general polynomial weights
$V_{N,\beta}(x)=\kappa_{2m}x^{2m}+\cdots$, $\kappa_{2m}>0$.

It turns out that not only can the statistics for OE's, UE's and
SE's be computed explicitly, but in the large $N$ limit the behavior
of these systems is universal in the sense described above, as
conjectured earlier by Dyson, Mehta, Wigner, and many others. It
works like this: Consider $N\times N$ matrices $M$ in a UE with
potential $V_{N,2}$. Let $K_N$ denote the finite rank operator with
kernel
\begin{equation}\label{E:7}
K_N(x,y)=\sum_{j=0}^{N-1} \varphi_j(x)\varphi_j(y),\qquad x,y\in\IR
\end{equation}
where
\begin{equation}\label{E:8}
\varphi_j(x)=p_j(x) e^{-\frac12 V_{N,2}(x)},\qquad j\geq0
\end{equation}
and
\begin{equation}\label{E:9}
p_j(x)=\gamma_j x^j+\cdots,\qquad j\geq0,\quad \gamma_j>0
\end{equation}
are the orthonormal polynomials with respect to the weight
$e^{-V_{N,2}(x)}\,dx$,
$$
\int_\IR p_j(x)p_k(x) e^{-V_{N,2}(x)}\,dx=\delta_{jk},\quad
j,k\geq0.
$$
Then the $m$-point correlation functions
\begin{equation*}
R_m(\lambda_1,\dots,\lambda_m)\equiv \frac{N!}{(N-m)!}
\int\!\!\!\cdots\!\!\!\int \hat P_{N,2}(\lambda_1,\dots,\lambda_N)
\,d\lambda_{m+1}\cdots d\lambda_N
\end{equation*}
can be expressed in terms of $K_N$ as follows:
\begin{equation}\label{E:10}
R_m(\lambda_1,\dots,\lambda_m)=\det\bigl(K_N(\lambda_i,\lambda_j)\bigr)_{1\leq
i,j\leq m}\,.
\end{equation}
A simple computation for the 1-point and 2-point functions,
$R_1(\lambda)$ and $R_2(\lambda_1,\lambda_2)$, shows that
\begin{equation}\label{E:11}
\Exp\bigl(\#\{\lambda_i\in B\}\bigr)=\int_B R_1(\lambda)\,d\lambda
\end{equation}
for any Borel set $B\subset\IR$, and
\begin{equation}\label{E:12}
\Exp\bigl(\#\{\text{ordered pairs}\,(i,j),\,i\neq
j\,:\,(\lambda_i,\lambda_j)\in\Delta\}\bigr) =\int\!\!\int_\Delta
R_2(\lambda_1,\lambda_2)\,d\lambda_1d\lambda_2
\end{equation}
for any Borel set $\Delta\subset\IR^2$.

It follows in particular from \eqref{E:11} that, for an energy $E$,
$R_1(E)=K_N(E,E)$ is the density of the expected number of
eigenvalues in a neighborhood of $E$, and hence, by the standard
procedure, one should take the scaling factor $\gamma_\lambda$ in
\eqref{E:5} to be $K_N(E,E)$. For energies $E$ in the bulk of the
spectrum, one finds for a broad class of potentials $V_{N,2}$ (see
\cite{DKMVZ1} and \cite{DKMVZ2}) that, in the scaling limit dictated
by $K_N(E,E)$, $K_N(\lambda,\lambda')$ takes on a universal form
\begin{equation}\label{E:12.1}
\lim_{N\to\infty} \frac{1}{K_N(E,E)}\,
K_N\left(E+\frac{x}{K_N(E,E)},E+\frac{y}{K_N(E,E)}\right)=K_\infty(x-y)
\end{equation}
where $x,y\in\IR$ and $K_\infty$ is the so-called
\textit{sine-kernel},
\begin{equation}\label{E:13}
K_\infty(u)=\frac{\sin(\pi u)}{\pi u}.
\end{equation}
Inserting this information into \eqref{E:10} we see that the scaling
limit for $R_m$ is universal for each $m\geq2$, and in particular for
$m=2$, we have for $x,y\in\IR$
\begin{equation}\label{E:14}
\begin{aligned}
\lim_{N\to\infty} \frac{1}{\bigl(K_N(E,E)\bigr)^2}\,
&R_2\left(E+\frac{x}{K_N(E,E)},E+\frac{y}{K_N(E,E)}\right)\\
&=\det \left(\begin{matrix}
K_\infty(0) & K_\infty(x-y)\\
K_\infty(x-y) & K_\infty(0)
\end{matrix}\right)\\
&=1-\left(\frac{\sin\pi(x-y)}{\pi(x-y)}\right)^2.
\end{aligned}
\end{equation}

For a Borel set $B\subset\IR$, let
$n_B=\#\{\lambda_j\,:\,\lambda_j\in B\}$ and let
\begin{equation}\label{E:14.1}
V_B=\Exp \bigl(n_B-\Exp (n_B)\bigr)^2
\end{equation}
denote the number variance in $B$. A simple computation again shows
that
$$
V_B=\int_B R_1(x)\,dx +\int\!\!\int_{B\times B} R_2(x,y)\,dxdy -
\left(\int_B R_1(x)\,dx\right)^2
$$
For an energy $E$ in the bulk of the spectrum as above, set
$$
B_N(s)=\left(E-\frac{s}{2K_N(E,E)},
E+\frac{s}{2K_N(E,E)}\right),\qquad s>0.
$$
For such $B$, $V_B$ is the number variance for an interval about $E$
of scaled size $s$. Recalling that $K_N(E,E)=R_1(E)$, and using
\eqref{E:14}, we find as $N\to\infty$
\begin{equation}\label{E:14.2}
\lim_{N\to\infty} V_{B_N(s)}= \frac{1}{\pi^2}\int_0^{2\pi s}
\frac{1-\cos u}{u}\,du + \frac{2s}{\pi}\int_{\pi s}^\infty
\left(\frac{\sin u}{u}\right)^2\,du.
\end{equation}
For large $s$, the right-hand side has the form (see \cite{Meh})
\begin{equation}\label{E:14.3}
\frac{1}{\pi^2} \bigl(\log(2\pi
s)+\gamma+1\bigr)+O\Bigl(\frac1s\Bigr)
\end{equation}
where $\gamma$ is Euler's constant.

For $\theta>0$, the so-called \textit{gap probability}
\begin{equation}\label{E:15}
G_{N,2}(\theta)=\text{Prob}\bigl(M\,:\, M\,\text{has no eigenvalues
in}\,(E-\theta,E+\theta)\bigr)
\end{equation}
is given by (see \cite{Meh}, and also \cite{Dei1})
\begin{equation}\label{E:16}
G_{N,2}(\theta)=\det\bigl(1-K_N\!\!\upharpoonright_{L^2(E-\theta,E+\theta)}\bigr)
\end{equation}
where $K_N\!\!\upharpoonright_{L^2(E-\theta,E+\theta)}$ denotes the
operator with kernel \eqref{E:7} acting on $L^2(E-\theta,E+\theta)$.
In the bulk scaling limit, we find
\begin{equation}\label{E:17}
\lim_{N\to\infty}
G_{N,2}\left(\frac{x}{K_N(E,E)}\right)=\det\bigl(1-K_\infty\!\!\upharpoonright_{L^2(E-\theta,E+\theta)}\bigr),
\qquad x\in\IR.
\end{equation}
In terms of the scaled eigenvalues
$\tilde\lambda_j=K_N(E,E)\cdot(\lambda_j-E)$, this means that for
$x>0$
\begin{equation}\label{E:17+}
\lim_{N\to\infty}\text{Prob}\bigl(M\,:\,\tilde\lambda_j\notin(-x,x),\,1\leq
j\leq N\bigr)
=\det\bigl(1-K_\infty\!\!\upharpoonright_{L^2(-x,x)}\bigr).
\end{equation}

Now consider a point $E$, say $E=0$, where
$K_N(E,E)=K_N(0,0)\to\infty$ as $N\to\infty$. This is true, in
particular, if
\begin{equation}\label{E:18}
V_{N,2}(x)=\kappa_mx^{2m}+\cdots,\qquad \kappa_m>0,\,m\geq1,
\end{equation}
and so for $V_{N,2}(x)=x^2$ (GUE). For such $V_{N,2}$'s,
we have $K_N(0,0)\sim N^{1-\frac{1}{2m}}$ (see \cite{DKMVZ1}). Let
$t_N>0$ be such that
\begin{equation}\label{E:19}
t_N\to\infty,\qquad \frac{t_N}{K_N(0,0)}\to0.
\end{equation}
Then
\begin{equation}\label{E:20}
\hat N\equiv
\Exp\left(\#\Bigl\{|\lambda_j|\leq\frac{t_N}{K_N(0,0)}\Bigr\}\right)
=\int_{-\frac{t_N}{K_N(0,0)}}^\frac{t_N}{K_N(0,0)}
K_N(\lambda,\lambda)\,d\lambda\sim2t_N\to\infty.
\end{equation}
For $a<b$, define the Borel set $\Delta_N\subset\IR^2$ by
\begin{equation}\label{E:21}
\Delta_N=\left\{(x,y)\,:\,\frac{a}{K_N(0,0)}<x-y<\frac{b}{K_N(0,0)}\,\,\text{and}\,\,
|x|,|y|\leq\frac{t_N}{K_N(0,0)}\right\}.
\end{equation}
Then we have by \eqref{E:12} and \eqref{E:14}, as $N\to\infty$,
\begin{align*}
&\frac{1}{\hat N} \Exp\bigl(\#\{\text{ordered pairs}\,(i,j),\,i\neq
j\,:\,
(\lambda_i,\lambda_j)\in\Delta_N\}\bigr)\\
&=\frac{1}{\hat N} \int\!\!\int_{\Delta_N} R_2(\lambda_1,\lambda_2)\,d\lambda_1d\lambda_2\\
&=\frac{1}{\hat N} \int\!\!\int_{\{(s,t):\,a<s-t<b,\,|s|,|t|<t_N\}}
\frac{1}{\bigl(K_N(0,0)\bigr)^2} R_2\left(\frac{s}{K_N(0,0)},\frac{t}{K_N(0,0)}\right)\,ds\,dt\\
&\sim\frac{1}{\hat N}
\int\!\!\int_{\{(s,t):\,a<s-t<b,\,|s|,|t|<t_N\}}
\left(1-\Bigl(\frac{\sin\pi(s-t)}{\pi(s-t)}\Bigr)^2\right)\,ds\,dt\\
&\sim \frac{2t_N}{\hat N}\int_a^b \left(1-\Bigl(\frac{\sin\pi r}{\pi r}\Bigr)^2\right)\,dr\\
&\sim\int_a^b \left(1-\Bigl(\frac{\sin\pi r}{\pi
r}\Bigr)^2\right)\,dr,\qquad\qquad\text{by \eqref{E:20}}.
\end{align*}
Thus, if $\tilde \lambda_j\equiv K_N(0,0)\lambda_j$ are, again, the
scaled eigenvalues, then for $t_N$ as in \eqref{E:19}
\begin{equation}\label{E:22}
\begin{aligned}
\lim_{N\to\infty} \frac{1}{\hat N} \Exp\bigl(\#\{\text{ordered
pairs}\, &(i,j), i\neq j\,:\, a<\tilde\lambda_i
-\tilde\lambda_j<b,\,|\tilde\lambda_i|,|\tilde\lambda_j|\leq t_N\}\bigr)\\
&=\int_a^b \left(1-\Bigl(\frac{\sin\pi r}{\pi r}\Bigr)^2\right)\,dr.
\end{aligned}
\end{equation}

Another quantity of interest is the spacing distribution of the
eigenvalues $\lambda_1\leq\lambda_2\leq\cdots\leq\lambda_N$ of a
random $N\times N$ matrix as $N\to\infty$. More precisely, for
$s>0$, we want to compute
$$
\Exp\left(\frac{\#\{1\leq j\leq N-1\,:\,
\lambda_{j+1}-\lambda_{j}\leq s\}}{N}\right)
$$
as $N$ becomes large. If we again restrict our attention to
eigenvalues in a neighborhood of a bulk energy $E=0$, say, then the
eigenvalue spacing distribution exhibits universal behavior for UE's
as $N\to\infty$. We have in particular the following result of
Gaudin (see \cite{Meh}, and also \cite{Dei1}): With $t_N$, $\hat N$
and $\tilde x_j=K_N(0,0)x_j$ as above,
\begin{equation}\label{E:23}
\begin{aligned}
&\lim_{N\to\infty} \Exp\left(\frac{\#\{1\leq j\leq N-1\,:\,
\tilde\lambda_{j+1}-\tilde\lambda_{j}\leq s,\, |\tilde\lambda_j|\leq t_N\}}{\hat N}\right)\\
&=\lim_{N\to\infty} \text{Prob}(\text{at least one
eigenvalue}\,\tilde\lambda_j\,
\text{in}\,(0,s]\,|\,\text{eigenvalue at 0})\\
&=\int_0^s p(u)\,du
\end{aligned}
\end{equation}
where
\begin{equation}\label{E:24}
p(u)=\frac{d^2}{du^2}\left(\det\bigl(1-K_\infty\!\!\upharpoonright_{L^2(0,u)}\bigr)\right).
\end{equation}

At the upper spectral edge $E=\lambda_\text{max}$, one again finds
universal behavior for UE's with potentials $V_{N,2}$, in
particular, of the form \eqref{E:18} above. For such $V_{N,2}$'s
there exist constants $z_N^{(2)},s_N^{(2)}$ such that for $t\in\IR$
(see \cite{DG2} and the notes therein)
\begin{equation}\label{E:25}
\lim_{N\to\infty}
\text{Prob}\left(M\,:\,\frac{\lambda_\text{max}-z_N^{(2)}}{s_N^{(2)}}\leq
t\right) =\det\bigl(1-\mathcal
A\!\upharpoonright_{L^2(t,\infty)}\bigr).
\end{equation}
Here $\mathcal A$ is the so-called \textit{Airy operator} with
kernel
\begin{equation}\label{E:26}
\mathcal A(x,y)=\frac{Ai(x)Ai'(y)-Ai'(x)Ai(y)}{x-y},
\end{equation}
where $Ai(x)$ is the classical Airy function. For GUE, where
$V_{N,2}(x)=x^2$, one has $z_N^{(2)}=\sqrt{2N}$ and
$s_N^{(2)}=2^{-\frac12}N^{-\frac16}$ (see Forrester \cite{F} and the
seminal work of Tracy and Widom \cite{TW2}).

It turns out that
$\det\bigl(1-K_\infty\!\!\upharpoonright_{L^2(-x,x)}\bigr)$ in
\eqref{E:17} and $\det\bigl(1-\mathcal
A\!\!\upharpoonright_{L^2(t,\infty)}\bigr)$ can be expressed in
terms of solutions of the Painlev\'e V and Painlev\'e II equations
respectively. The first is a celebrated result of Jimbo, Miwa,
M\^{o}ri, and Sato \cite{JMMS}, and the second is an equally
celebrated result of Tracy and Widom \cite{TW2}. In particular for
edge scaling we find
\begin{equation}\label{E:27}
\lim_{N\to\infty}\text{Prob}\left(M\,:\,\frac{\lambda_\text{max}-z_N^{(2)}}{s_N^{(2)}}\leq
t\right)=F_{\beta=2}(t)
\end{equation}
where
\begin{equation}\label{E:28}
F_{\beta=2}(t)=\det\bigl(1-\mathcal
A\!\upharpoonright_{L^2(t,\infty)}\bigr) =e^{-\int_t^\infty
(s-t)u^2(s)\,ds}
\end{equation}
and $u(s)$ is the (unique, global) Hastings-McLeod solution of the
Painlev\'e II equation
\begin{equation}\label{E:29}
u''(s)=2u^3(s)+su(s)
\end{equation}
such that
\begin{equation}\label{E:30}
u(s)\sim Ai(s)\qquad\qquad \text{as}\quad s\to+\infty.
\end{equation}
$F_2(t)=F_{\beta=2}(t)$ is called the \textit{Tracy-Widom
distribution} for $\beta=2$.

Finally we note that for OE's and SE's there are analogs for all the
above results \eqref{E:10}--\eqref{E:30}, and again one finds
universality in the scaling limits as $N\to\infty$ for potentials
$V_{N,\beta}$, $\beta=1,4$, of the form \eqref{E:18} above (see
\cite{DG1} and \cite{DG2} and the historical notes therein). We
note, in particular, the following results: for $V_{N,\beta}$ as
above, $\beta=1$ or 4, there exist constants
$z_N^{(\beta)},s_N^{(\beta)}$ such that
\begin{equation}\label{E:31}
\lim_{N\to\infty}\text{Prob}\left(M\,:\,\frac{\lambda_\text{max}(M)-z_N^{(\beta)}}{s_N^{(\beta)}}\leq
t\right)=F_{\beta}(t)
\end{equation}
where
\begin{equation}\label{E:32}
F_1(t)=\bigl(F_2(t)\bigr)^\frac12 e^{-\frac12\int_t^\infty u(s)\,ds}
\end{equation}
and
\begin{equation}\label{E:33}
F_4(t)=\bigl(F_2(t)\bigr)^\frac12\cdot \frac{e^{\frac12\int_t^\infty
u(s)\,ds}+e^{-\frac12\int_t^\infty u(s)\,ds}}{2}
\end{equation}
with $F_2(t)$ and $u(s)$ as above. $F_1(t)$ and $F_4(t)$ are called
the \textit{Tracy-Widom distributions} for $\beta=1$ and 4 respectively.

\section{The problems}\label{S:3}

In this section we consider seven problems. The first is
from physics and is included for historical reasons that will become
clear in Section~\ref{S:4} below; the remaining six problems are
from mathematics/mathematical physics.

\bigskip

\noindent \textbf{Problem 1}

Consider the scattering of neutrons off a heavy nucleus, say uranium
U$^{238}$. The scattering cross-section is plotted as a function of
the energy $E$ of the incoming neutrons, and one obtains a jagged
graph (see \cite{Por} and \cite{Meh}) with many hundreds of sharp
peaks $E_1<E_2<\cdots$ and valleys $E'_1<E'_2<\cdots$. If $E\sim
E_j$ for some $j$, the neutron is strongly repelled from the
nucleus, and if $E\sim E'_j$ for some $j$, then the neutron sails
through the nucleus, essentially unimpeded. The $E_j$'s are called
\textit{scattering resonances}. The challenge faced by physicists in
the late 40's and early 50's was to develop an effective model to
describe these resonances. One could of course write down a
Schr\"odinger-type equation for the scattering system, but because
of the high dimensionality of the problem there is clearly no hope
of solving the equation for the $E_j$'s either
analytically or numerically. However, as more experiments were done
on heavy nuclei, each with hundreds of $E_j$'s, a consensus began to
emerge that the ``correct'' theory of resonances was statistical,
and here Wigner led the way. Any effective theory would have to
incorporate two essential features present in the data, viz.
\begin{itemize}
  \item[\rm{(i)}] modulo certain natural symmetry considerations, all nuclei in the same symmetry class
     exhibited universal behavior
  \item[\rm{(ii)}] in all cases, the $E_j$'s exhibited \textit{repulsion}, or, more precisely, the probability
     that two $E_j$'s would be close together was small.
\end{itemize}

\smallskip

\noindent \textbf{Question 1.} What theory did Wigner propose for
the $E_j$'s?

\bigskip

\noindent \textbf{Problem 2}

Here we consider the work of H.~Montgomery \cite{Mon} in the early
1970's on the zeros of the Riemann zeta function
$\zeta(s)$. Assuming the Riemann hypothesis, Montgomery rescaled the
imaginary parts $\gamma_1\leq\gamma_2\leq\cdots$ of the (nontrivial) zeros
$\{\frac12+i\gamma_j\}$ of $\zeta(s)$,
\begin{equation}\label{E:34}
\gamma_j\to\tilde\gamma_j=\frac{\gamma_j\log\gamma_j}{2\pi}
\end{equation}
to have mean spacing 1 as $T\to\infty$, i.e.
$$
\lim_{T\to\infty} \frac{\#\{j\geq1\,:\,\tilde\gamma_j\leq T\}}{T}=1.
$$
For any $a<b$, he then computed the two-point correlation function
for the $\tilde\gamma_j$'s
$$
\#\{\text{ordered pairs}\,(j_1,j_2),j_1\neq j_2\,:\,1\leq
j_1,j_2\leq N,\tilde\gamma_{j_1}-\tilde\gamma_{j_2}\in(a,b)\}
$$
and showed, modulo certain technical restrictions, that
\begin{equation}\label{E:35}
R(a,b)\equiv \lim_{N\to\infty} \frac1N\#\{\text{ordered
pairs}\,(j_1,j_2),j_1\neq j_2\,:\,1\leq j_1,j_2\leq N,
\tilde\gamma_{j_1}-\tilde\gamma_{j_2}\in(a,b)\}
\end{equation}
exists and is given by a certain explicit formula.

\smallskip

\noindent \textbf{Question 2.} What formula did Montgomery obtain
for $R(a,b)$?

\bigskip

\noindent \textbf{Problem 3}

Consider the solitaire card game known as \textit{patience sorting}
(see \cite{AldDia} and \cite{Mal}). The game is played with $N$
cards, numbered 1,2,...,$N$ for convenience. The deck is shuffled
and the first card is placed face up on the table in front of the
dealer. If the next card is smaller than the card on the table, it
is placed face up on top of the card; if it is bigger, the card is
placed face up to the right of the first card, making a new pile. If
the third card in the pile is smaller than one of the cards on the
table, it is placed on top of that card; if it is smaller than both
cards, it is placed as far to the left as possible. If it is bigger
than than both cards, it is placed face up to the right of the
pile(s), making a new pile. One continues in this fashion until all
the cards are dealt out. Let $q_N$ denote the number of piles
obtained. Clearly $q_N$ depends on the particular shuffle $\pi\in S_N$, the
symmetric group on $N$ numbers, and we write $q_N=q_N(\pi)$.

For example, if $N=6$ and $\pi=3\,4\,1\,5\,6\,2$, where $3$ is the
top card, 4 is the next card and so on, then patience sorting
proceeds as follows:
\begin{equation*}
\begin{matrix}  \\3 \end{matrix}
\quad\qquad \begin{matrix}&  \\ 3& 4  \end{matrix} \quad\qquad
\begin{matrix} 1 &  \\ 3& 4  \end{matrix} \quad\qquad\begin{matrix}1
& &  \\ 3 & 4 & 5  \end{matrix} \quad\qquad \begin{matrix}1 & & & \\
3 & 4 & 5 & 6 \end{matrix} \quad\qquad \begin{matrix}1 & 2 & & \\ 3&
4 & 5 & 6 \end{matrix}
\end{equation*}
and $q_6(\pi)=4$.

\smallskip

\noindent \textbf{Question 3.} Equip $S_N$ with the uniform
distribution. If each card is of unit size, how big a table does one
typically need to play patience sorting with $N$ cards? Or, more
precisely, how does
\begin{equation}\label{E:36}
p_{n,N}=\text{Prob} \bigl(\pi\,:\,q_N(\pi)\leq n\bigr)
\end{equation}
behave as $N\to\infty$, $n\leq N$?

\bigskip

\noindent \textbf{Problem 4}

The city of Cuernavaca in Mexico (population about 500,000) has an
extensive bus system, but there is no municipal transit authority to control
the city transport. In particular there is no timetable, which gives rise
to Poisson-like phenomena, with bunching and long waits between
buses. Typically, the buses are owned by drivers as individual
entrepreneurs, and all too often a bus arrives at a stop just
as another bus is loading up. The driver then has to move on
to the next stop to find his fares. In order to remedy the situation
the drivers in Cuernavaca came up with a novel solution: they
introduced ``recorders'' at specific locations along the bus routes
in the city. The recorders kept track of when buses passed their
locations, and then sold this information to the next driver, who
could then speed up or slow down in order to optimize the distance
to the preceding bus. The upshot of this ingenious scheme is that
the drivers do not lose out on fares and the citizens of Cuernavaca
now have a reliable and regular bus service. In the late 1990's two
Czech physicists with interest in transportation problems,
M.~Krb\'alek and P.~\v{S}eba, heard about the buses in Cuernavaca
and went down to Mexico to investigate. For about a month they
studied the statistics of bus arrivals on Line 4 close to the city
center. In particular, they studied the bus spacing distribution,
and also the bus number variance measuring the fluctuations of the
total number of buses arriving at a fixed location during a time
interval $T$. Their findings are reported in \cite{KrbSeb}.

\smallskip

\noindent \textbf{Question 4.} What did Krb\'alek and \v{S}eba learn
about the statistics of the bus system in Cuernavaca?

\bigskip

\noindent \textbf{Problem 5}

In his investigation of wetting and melting phenomena in \cite{Fis},
Fisher introduced various ``vicious'' walker models. Here we will
consider the so-called \textit{random turns vicious walker model}.
In this model, the walks take place on the integer lattice $\ZZ$ and
initially the walkers are located at $0,1,2,...\,$. The rules for a
walk are as follows:
\begin{itemize}
  \item[(a)] at each  tick of the clock, precisely one walker makes a step to the left
  \item[(b)] no two walkers can occupy the same site (hence ``vicious walkers'')
\end{itemize}
For example, consider the following walk from time $t=0$ to time
$t=4$:
\begin{figure}[h]
\def\p{\hbox to 0mm{\hss+\hss}}
{ \footnotesize \setlength{\arraycolsep}{1ex}
\begin{equation*}
\begin{matrix}
\cdot & \cdot &\times & \cdot &\times &\times & \cdot &\times &\times \\
\cdot & \cdot &\times & \cdot &\times & \cdot &\times &\times &\times \\
\cdot & \cdot & \cdot &\times &\times & \cdot &\times &\times &\times \\
\cdot & \cdot & \cdot &\times & \cdot &\times &\times &\times &\times \\
\cdot & \cdot & \cdot & \cdot &\times &\times &\times &\times &\times \\
      &       &       &       &       &       &       &       &       \\
-4    &  -3   &   -2  &   -1  &  0    &   1   &  2    &   3   &   4   \\
\end{matrix}
\qquad\qquad
\begin{matrix}
t=4\\ t=3\\ t=2\\ t=1\\ t=0\\ \\ \\
\end{matrix}
\end{equation*}
\vskip -1ex } \caption{Random turns walk.}\label{F1}
\end{figure}
At $t=0$, clearly only the walker at 0 can move. At time $t=1$,
either the walker at -1 or at +1 can move, and so on. Let $d_N$ be
the distance traveled by the walker starting from 0. In the above
example, $d_4=2$. For any time $N$, there are clearly only a finite
number of possible walks of duration $t=N$. Suppose that all such
walks are equally likely.

\smallskip

\noindent \textbf{Question 5.} How does $d_N$ behave statistically
as $N\to\infty$?

\bigskip

\noindent \textbf{Problem 6}

\begin{figure}[h]
\begin{center}
\epsfig{file=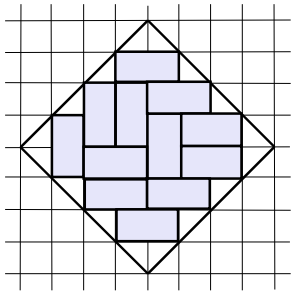,width=1.5in}
\end{center}
\caption{Aztec diamond for $n=3$.}\label{F:diamond}
\end{figure}

Consider tilings $\{T\}$ of the tilted square
$\mathbf{T}_n=\{(x,y)\,:\,|x|+|y|\leq n+1\}$ in $\IR^2$ by
horizontal and vertical dominos of length 2 and width 1.
For example, for $n=3$ we have the tiling $T$ of Figure~\ref{F:diamond}.
For each tiling the dominos must lie
strictly within $\mathbf{T}_n$. The tilings $T$ are called
\textit{Aztec diamonds} because the boundary of $T$ in $\{(x,y)\,:\,
y>0\}$, say, has the shape of a Mexican pyramid. It is a nontrivial
theorem (see \cite{Elk...}) that for any $n$, the number of domino
tilings of $\mathbf{T}_n$ is $2^{\frac{n(n+1)}{2}}$. Assume that all
tilings are equally likely.

\smallskip

\noindent \textbf{Question 6.} What does a typical tiling look like
as $n\to\infty$?

\smallskip

Finally we have

\bigskip

\noindent \textbf{Problem 7}

How long does it take to board an airplane? We consider the random
boarding strategy in \cite{BacBerSapSkiSto} under the following
simplifying assumptions:
\begin{itemize}
\item[(a)] there is only 1 seat per row
\item[(b)] the passengers are very thin compared to the distance between seats
\item[(c)] the passengers move very quickly between seats. The main delay in boarding
is the time - one unit - that it takes for the passengers to
organize their luggage and seat themselves once they arrive at their
assigned seats.
\end{itemize}
For the full problem with more than one seat per row, passengers who are not
``very thin'', etc., see \cite{BacBerSapSkiSto}, and also the discussion of the boarding problem
in Section~\ref{S:4} below.

The passengers enter the airplane through a door in front and the
seats are numbered 1,2,..., $N$, with seat 1 closest to the door.
How does boarding proceed? Consider, for example, the case $N=6$.
There are 6 passengers, each with a seating card 1,2,...,6. At the
call to board, the passengers line up randomly at the gate. Suppose
for definitiveness that the order in the line is given by
\begin{equation}\label{E:37}
\pi:\quad 3\,4\,1\,5\,6\,2
\end{equation}
with 3 nearest the gate. Now 3 can proceed to his'r seat, but 4 is
blocked and must wait behind 3 while s'he puts his'r bag up into the
overhead bin. However, at the same time, 1 can proceed to his'r
seat, but 5,6, and 2 are blocked. At the end of one unit of time, 3
and 1 sit down, and 4 and 2 can proceed to their seats, but 5 and 6
are blocked behind 4. After one more unit of time 4 and 2 sit down,
and 5 can proceed to his'r seat, but 6 is blocked. At the end of one
more unit of time 5 sits down, and finally 6 can move to his'r
seat. Thus for $\pi$ as above, it takes 4 units of time to board.
Let $b_N=b_N(\pi)$ denote the boarding time for any $\pi\in S_N$,
and assume that the $\pi$'s are uniformly distributed.

\smallskip

\noindent \textbf{Question 7.} How does $b_N(\pi)$ behave
statistically as $N\to\infty$?

\section{Solutions and explanations}\label{S:4}

As indicated in the Introduction, the remarkable fact of the matter
is that all seven problems in Section~\ref{S:3} are modeled by RMT.

\bigskip

\noindent \textbf{Problem 1.} (neutron scattering)

At some point in the mid-1950's, in a striking development, Wigner
suggested that the statistics of the neutron scattering resonances
was governed by GOE\footnote{Here we must restrict the data to
scattering for situations where the nuclear forces are time-reversal
invariant. If not, the statistics of the scattering resonances
should be governed by GUE.} (and hence, by universality \cite{DG1},
by all OE's). And indeed, if one scales real scattering data for a
variety of nuclei according to the standard procedure and then
evaluates, in particular, the nearest neighbor distribution, one finds
remarkable agreement with the OE analog of the spacing distribution \eqref{E:23},\eqref{E:24}.

It is interesting, and informative, to trace the development of
ideas that led Wigner to his suggestion (see \cite{Wig1},
\cite{Wig2}, \cite{Wig3}; all three papers are reproduced in
\cite{Por}). In these papers, Wigner is guided by the fact that any
model for the resonances would have to satisfy the constraints of
universality and repulsion, (i) and (ii) respectively, in the
description of Problem~1. In \cite{Wig2} he recalls a paper that he
had written with von Neumann in 1929 in which
they showed, in particular, that in the
$\frac{n(n+1)}{2}$-dimensional space of real $n\times n$ symmetric
matrices, the matrices with double eigenvalues form a set of
codimension 2. For example, if a $2\times2$ real symmetric matrix
has double eigenvalues, then it must be a multiple of the identity
and hence it lies in a set of dimension 1 in $\IR^3$. It follows that if one
equips the space of real, symmetric matrices with a probability
measure with a smooth density, the probability of a matrix $M$
having equal eigenvalues would be zero and the eigenvalues
$\lambda_1,\ldots,\lambda_n$ of $M$ would comprise a random set with
repulsion built in. So Wigner had a model, or more precisely, a
class of models, which satisfied constraint (ii). But why choose
GOE? This is where the universality constraint (i) comes into play.
We quote from \cite{Wig3}\footnote{In the quotation that follows, ``Figure 1'' portrays a level
spacing distribution, the ``Wishart distribution'' is the
statisticians' name for GOE, and ``Gatlinburg'' is \cite{Wig1}.}:
``Let me say only one more word. It is very likely that the curve in
Figure~1 is a universal function. In other words, it doesn't depend
on the details of the model with which you are working. There is one
particular model in which the probability of the energy levels can
be written down exactly. I mentioned this distribution already in
Gatlinburg. It is called the Wishart distribution. Consider a
set... .'' So in this way Wigner introduced GOE into theoretical
physics: It provided a model with repulsion (and time-reversal)
built in. Furthermore, the energy level distribution could be
computed explicitly. By universality, it should do the trick!

As remarkable as these developments were, even the most prophetic
observer could not have predicted that, a few years down the
line, these developments would make themselves felt within pure
mathematics.

\bigskip

\noindent \textbf{Problem 2.} (Riemann zeta function)

Soon after completing his work on the scaling limit \eqref{E:35} of
the two-point correlation function for the zeros of zeta, Montgomery
was visiting the Institute for Advanced Study in Princeton and it
was suggested that he show his result to Dyson. What happened is a
celebrated, and oft repeated, story in the lore of the Institute:
before Montgomery could describe his hard won result to Dyson, Dyson
took out a pen, wrote down a formula, and asked Montgomery ``And did
you get this?''
\begin{equation}\label{E:38}
R(a,b)=\int_a^b 1-\left(\frac{\sin(\pi r)}{\pi r}\right)^2\,dr
\end{equation}
Montgomery was stunned: this was exactly the formula he had
obtained. Dyson explained: ``If the zeros of the zeta function
behaved like the eigenvalues of a random GUE matrix, then
\eqref{E:38} would be exactly the formula for the two-point
correlation function!'' (See \eqref{E:22} above.)

More precisely, what Montgomery actually proved was that
\begin{equation}\label{E:39}
\lim_{N\to\infty} \sum_{1\leq i\neq j\leq N}
f(\tilde\gamma_i-\tilde\gamma_j) =\int_\IR f(r)
\left(1-\Bigl(\frac{\sin\pi r}{\pi r}\Bigr)^2\right)\,dr
\end{equation}
for any rapidly decaying function $f$ whose Fourier transform $\hat
f(\xi)$ is supported in the interval $|\xi|<2$. Of course, if one could
prove \eqref{E:39} for all smooth, rapidly decaying functions, one
would recover the full result \eqref{E:38}. Nevertheless, in an
impressive series of numerical computations starting in the 1980's,
Odlyzko verified \eqref{E:38} to extraordinary accuracy (see
\cite{Odl1}, \cite{Odl2}, and the references therein). In his
computations, Odlyzko also considered GUE behavior for other
statistics for the $\tilde\gamma_j$'s, such as the nearest neighbor
spacing, verifying in particular the relationship
\begin{equation}\label{E:40}
\lim_{N\to\infty}
\frac1N\#\{1\leq j\leq
N-1\,:\,\tilde\gamma_{j+1}-\tilde\gamma_j\leq s\} =\int_0^s
p(u)\,du,\qquad s>0,
\end{equation}
(cf \eqref{E:23}, \eqref{E:24}) to high accuracy.

The relationship between the zeros of the zeta function and random
matrix theory first discovered by Montgomery has been taken up with
great virtuosity by many researchers in analytic number theory, with
Rudnick and Sarnak \cite{RudSar}, and then Katz and Sarnak
\cite{KatSar}, leading the way. GUE behavior for the zeros of quite
general automorphic L-functions over $\IQ$, as well as for a wide
class of zeta and L-functions over finite fields, has now been
established (modulo technicalities as in \eqref{E:39} above in the
number field case). Another major development has been the discovery
of a relationship between random polynomials whose roots are given
by the eigenvalues of a matrix from some random ensemble, and the
moments of the L-functions on the critical line $\Re z=\frac12$ (see
\cite{KeaSna1}, \cite{KeaSna2}). The discovery of Montgomery/Odlyzko
counts as one of the major developments in analytic number theory in
many, many years.

\bigskip

\noindent \textbf{Problem 3.} (Patience sorting)

In 1999 Baik, Deift and Johansson \cite{BDJ1} proved the following
result for $q_N(\pi)$, the number of piles obtained in patience
sorting starting from a shuffle $\pi$ of $N$ cards. Let
$\chi_N=\frac{q_N-2\sqrt{N}}{N^{^{1/6}}}$. Then
\begin{equation}\label{E:41}
\lim_{N\to\infty} \text{Prob} \bigl(\chi_N\leq t\bigr)=F_2(t)
\end{equation} where $F_2$ is the Tracy-Widom distribution \eqref{E:27}, \eqref{E:28} for
$\beta=2$. Thus the number of piles, suitably centered and scaled,
behaves statistically like the largest eigenvalue of a GUE matrix.
In addition, the authors proved convergence of moments. For any
$m=1,2,\dots$,
\begin{equation}\label{E:42}
\lim_{N\to\infty} \Exp\bigl(\chi_N^m\bigr)=\Exp\bigl(\chi^m\bigr)
\end{equation}
where $\chi$ is any random variable with distribution $F_2$. In
particular, for $m=1,2$ one obtains
\begin{equation}\label{E:43}
\lim_{N\to\infty} \frac{\Exp(q_N)-2\sqrt{N}}{N^{^{1/6}}}=\int_\IR
t\,dF_2(t)
\end{equation}
and
\begin{equation}\label{E:44}
\lim_{N\to\infty} \frac{\mathbb{V}\text{ar}(q_N)}{N^{^{1/3}}} =\int_\IR
t^2\,dF_2(t) -\left(\int_\IR t\,dF_2(t)\right)^2.
\end{equation}
Numerical evaluation shows that the constants on the right-hand side
of \eqref{E:43} and \eqref{E:44} are given by -1.7711 and 0.8132,
respectively. Thus, as $N\to\infty$,
$$
\Exp(q_N)\sim 2\sqrt{N}-1.7711\cdot N^{1/6}
$$
so that for a deck of $N=52$ cards, one needs a table of size about
12 units on average to play the game.

Patience sorting is closely related to the problem of
\textit{longest increasing subsequences} for permutations $\pi\in
S_N$. Recall that we say that $\pi(i_1),\ldots,\pi(i_k)$ is an
increasing subsequence in $\pi$ of length $k$ if
$i_1<i_2<\cdots<i_k$ and $\pi(i_1)<\pi(i_2)<\cdots<\pi(i_k)$. Let
$l_N(\pi)$ be the length of the longest increasing subsequence in
$\pi$. For example, if $N=6$ and $\pi=3\,4\,1\,5\,6\,2$ we see that
$3\,4\,5\,6$ is a longest increasing subsequence for $\pi$ and hence
$l_6(\pi)=4$. Comparing with the introduction to Question~3, we see
that $l_6(\pi)=q_6(\pi)$. This is no accident: for any $\pi\in S_N$,
we always have $l_N(\pi)=q_N(\pi)$ (see, e.g. \cite{AldDia}), and
hence we learn from \eqref{E:41} that the length $l_N$ of the
longest increasing subsequence behaves statistically like the
largest eigenvalue of a GUE matrix as $N\to\infty$.
The relation $l_N(\pi)=q_N(\pi)$ and \eqref{E:43} imply in particular
that
\begin{equation}\label{E:49+}
\lim_{N\to\infty} \frac{\Exp(l_N)}{N^{^{1/2}}}=2.
\end{equation}
The claim that the limit in \eqref{E:49+} exists, and equals 2, is known as ``Ulam's problem'' and has a long
history (see \cite{BDJ1}).
In another direction, uniform distribution on $S_N$ pushes forward under the
Robinson-Schensted correspondence (see, e.g. \cite{Sag}) to
so-called \textit{Plancherel measure} on Young diagrams of size $N$.
Young diagrams are parameterized by partitions $\mu \vdash N$,
$\{\mu=(\mu_1,\mu_2,\ldots,\mu_l)\,:\,\mu_1\geq\mu_2\geq\cdots
\geq\mu_l\geq1,\,\sum_{i=1}^l \mu_i=N\}$, where $\mu_i$ is the
number of boxes in the $i^\text{th}$ row, and it turns out that
under the correspondence we have
\begin{equation}\label{E:45}
\text{Prob}\bigl(\pi\,:\, l_N(\pi)\leq n\bigr)
=\text{Prob}\bigl(\mu\vdash N\,:\,\mu_1\leq n\bigr).
\end{equation}
Consequently, the number of boxes in the first row of
Plancherel-random Young diagrams behaves statistically, as
$N\to\infty$, like the largest eigenvalue of a GUE matrix. In
\cite{BDJ1} the authors conjectured that the number of boxes in the
first $k$ rows of a Young diagram should behave statistically as
$N\to\infty$ like the top $k$ eigenvalues
$\lambda_N\geq\lambda_{N-1}\geq\cdots\geq\lambda_{N-k+1}$ of a GUE
matrix. This conjecture was proved for the $2^\text{nd}$ row in
\cite{BDJ2}. For general $k$, the conjecture was proved, with
convergence in joint distribution, in three separate papers in rapid
succession (\cite{Oko}, \cite{BorOkoOls}, \cite{Joh1}), using very
different methods. The proof in \cite{Oko} relies on an interplay between
maps on surfaces and ramified coverings of the sphere; the proof in \cite{BorOkoOls}
is based on the analysis of specific characters on $S(\infty)$, the infinite symmetric
group defined as the inductive limit of the finite symmetric groups $S_N$
under the embeddings $S_N\hookrightarrow S_{N+1}$; and the proof in \cite{Joh1} utilizes
certain discrete orthogonal polynomial ensembles arising in combinatorial probability.

One can consider the statistics of $l_N(\pi)$ for $\pi$ restricted
to certain distinguished subsets of $S_N$ (see \cite{BR}). Amongst
the many results in \cite{BR} relating combinatorics and random
matrix theory, we mention the following. Let
$S_N^\text{(inv)}=\{\pi\in S_N\,:\, \pi^2=id\}$ be the set of
involutions in $S_N$. Then, under the Robinson-Schensted
correspondence, uniform distribution on $S_N^\text{(inv)}$ pushes
forward to a new measure on Young diagrams, different from the
Plancherel measure. Denote this measure by
$\text{Prob}^\text{(inv)}$, and in place of \eqref{E:45} we have
$$
\text{Prob}\bigl(\pi\in S_N^\text{(inv)}\,:\, l_N(\pi)\leq n\bigr)
=\text{Prob}^\text{(inv)}\bigl(\mu\vdash N\,:\, \mu_1\leq n\bigr).
$$
In \cite{BR} the authors show that
\begin{equation}\label{E:46}
\begin{aligned}
&\quad\lim_{N\to\infty} \text{Prob}\Bigl(\pi\in S_N^\text{(inv)}\,:\,\frac{l_N-2\sqrt{N}}{N^{^{1/6}}}\leq x\Bigr)\\
&=\lim_{N\to\infty} \text{Prob}^\text{(inv)}\Bigl(\mu\vdash N\,:\,\frac{\mu_1-2\sqrt{N}}{N^{^{1/6}}}\leq x\Bigr)\\
&=F_{\beta=1}(x)
\end{aligned}
\end{equation}
and for the second row of $\mu$
\begin{equation}\label{E:47}
\lim_{N\to\infty}\text{Prob}^\text{(inv)}\Bigl(\mu\vdash N\,:\,
\frac{\mu_2-2\sqrt{N}}{N^{^{1/6}}}\leq x\Bigr) =F_{\beta=4}(x).
\end{equation}
Here $F_{\beta=1}$ and $F_{\beta=4}$ are the Tracy-Widom
distributions for the largest eigenvalue of the GOE and GSE
ensemble, respectively (see \eqref{E:31},\eqref{E:32},\eqref{E:33}).
Thus all three of the basic ensembles $\beta=1,2$ and 4 show up in
the analysis of the (general) increasing subsequence problem.

A problem which is closely related to the longest increasing
subsequence problem is the random word problem. In \cite{TW3} the
authors consider words $\{\omega\}$ of length $N$ in an alphabet of
$k$ letters, i.e. maps
$\omega\,:\,\{1,2,\ldots,N\}\to\{1,2,\ldots,k\}$. One says that
$\omega(i_1),\ldots,\omega(i_j)$ is a weakly increasing subsequence
in $\omega$ of length $j$ if $i_1<i_2<\cdots<i_j$ and
$\omega(i_1)\leq\omega(i_2)\leq\cdots\leq\omega(i_j)$. Let
$l_N^{^\text{wk}}(\omega)$ denote the length of the longest weakly
increasing subsequence in $\omega$. Assuming that all words are
equally likely, Tracy and Widom in \cite{TW3} proved that
\begin{equation}\label{E:48}
\lim_{N\to\infty}
\text{Prob}\Bigl(\omega\,:\,\frac{l_N^{^\text{wk}}(\omega)-\tfrac{N}{k}}{\sqrt{\tfrac{2N}{k}}}\leq
s\Bigr) =\gamma_k\int_{\mathcal L_s} e^{-\sum_{i=1}^k x_i^2}
\prod_{1\leq i<j\leq k} (x_i-x_j)^2\,dx_1\cdots dx_{k-1}
\end{equation}
where
\begin{equation}\label{E:49}
\mathcal L_s=\{(x_1,\ldots,x_k)\,:\, \max_{1\leq i\leq k} x_i\leq s,\,x_1+\cdots+x_k=0\}
\end{equation}
and
\begin{equation}\label{E:50}
\gamma_k=\frac{\sqrt{k}\,2^{^{\tfrac{k^2-1}{k}}}}{\Bigl(\prod_{i=1}^k
i!\Bigr)\,(2\pi)^{^{\tfrac{k-1}{2}}}}\,.
\end{equation}
It is easy to see that the right-hand side of \eqref{E:48} is just
the distribution function for the largest eigenvalue of a $k\times k$ GUE
matrix conditioned to have trace zero.

Consider the representation of the number $\pi$, say, in any basis
$b$,
\begin{equation}\label{E:51}
\pi=0.a_1a_2a_3\ldots\times b^q,\qquad q\in\ZZ.
\end{equation}
It has long been believed that in some natural asymptotic sense the
digits $a_1, a_2, a_3,\ldots$ are independent and identically
distributed, with uniform distribution on $\{0,1,\ldots,b-1\}$. In an
attempt to formalize this notion, E.~Borel  (1909) introduced the
idea of normality (see \cite{Wag}): A real number $x$ is \textit{normal}
if for any base $b$, any $m\geq 1$, and any $m$-string $s$,
\begin{equation}\label{E:52}
\lim_{n\to\infty} \frac{\#\{\text{occurrences of $s$ in the first
$n$ base-$b$ digits of $x$}\}}{n} =b^{-m}.
\end{equation}
While it is known that non-normal numbers form a set of Lebesgue
measure zero, and all numerical evidence confirms \eqref{E:52} to
high order, no explicit examples of normal numbers are known.

Relation \eqref{E:48} suggests a new way to test for asymptotic
randomness, as follows. Consider the first $LN$ base-$b$ digits
$a_1a_2\ldots a_{LN}$ of a given number $x$, where $L$ and $N$ are
``large''. Partition these digits into $L$ words
$\omega_j=a_{(j-1)N+1}\cdots a_{jN}$, $1\leq j\leq L$, each of
length $N$. For each $\omega_j$ compute
$l_N^{^\text{wk}}(\omega_j)$. Then if the digits $\{a_j\}$ of $x$
are asymptotically random, we could expect that as $L,N\to\infty$,
the empirical distribution
$$
\frac1L \#\Biggl\{1\leq j\leq L\,:\,
\frac{l_N^{^\text{wk}}(\omega_j)-\tfrac{N}{b}}{\sqrt{\tfrac{2N}{b}}}\leq
s \Biggr\}
$$
is close to the conditional GUE distribution on the right-hand side
of \eqref{E:48}. Preliminary calculations in \cite{DeiWit} for
$x=\pi$ and $b=2$ show that for $L,N$ ``large'' the empirical distribution is indeed
close to the right-hand side of \eqref{E:48} with high accuracy. The
work is in progress.

\bigskip

\noindent \textbf{Problem 4.} (Bus problem in Cuernavaca)

Krb\'alek and \v{S}eba found that both
the bus spacing distribution and the number variance are well
modeled by GUE, \eqref{E:23}, \eqref{E:24} and \eqref{E:14.2}
respectively (see Figures~2 and 3 in \cite{KrbSeb}). In order to
provide a plausible explanation of the observations in
\cite{KrbSeb}, the authors in \cite{BaiBorDeiSui} introduced a
microscopic model for the bus line that leads simply and directly to GUE.

The main features of the bus system in Cuernavaca are
\begin{itemize}
\item[(a)] the stop-start nature of the motion of the buses,
\item[(b)] the ``repulsion'' of the buses due to the presence of recorders.
\end{itemize}
To capture these features, the authors in \cite{BaiBorDeiSui}
introduced a model for the buses consisting of $n$(= \# of buses)
independent, rate 1 Poisson processes moving from the bus depot at
time $t=0$ to the final terminus at
time $T$, and conditioned not to intersect for $0\leq t\leq T$. The
authors then showed that at any observation point $x$ along the
route of length $N>n$, the probability distribution for the
(rescaled) arrival times of the buses,
$y_j=\frac{2t_j}{T}-1\in[-1,1]$, $1\leq j\leq n$, is given by
\begin{equation}\label{E:53}
\text{const.}\prod_{j=1}^n w_{_J}(y_j) \prod_{1\leq i<j\leq n}
(y_i-y_j)^2\,dy_1\cdots dy_n
\end{equation}
where
\begin{equation}\label{E:54}
w_{_J}(y)=(1+y)^{x-1}(1-y)^{N-x-n+1},\qquad -1<y<1.
\end{equation}
Formula \eqref{E:53} is precisely the eigenvalue distribution for
the so-called Jacobi Unitary Ensemble (cf. \eqref{E:4} with
$e^{-V_{N,2}(y)}=w_J(y)=$ weight for Jacobi polynomials on
$[-1,1]$). In the appropriate scaling limit, GUE then emerges by
universality. The authors also compute the distributions of the
positions $x_1,\ldots,x_n$ of the buses at any time $t\in(0,T)$.
Again the statistics of the $x_j$'s are described by a Unitary
Ensemble, but now $w_{_J}$ is \eqref{E:53} is replaced by the weight for the Krawtchouk polynomials: by
universality, GUE again emerges in the appropriate scaling limit.

In an intriguing recent paper, Abul-Magd~\cite{Abu} noted that
drivers have a tendency ``to park their cars near to each other and
at the same time keep a distance sufficient for manoeuvring.'' He
then analyzed data measuring the gaps between parked cars on four
streets in central London and showed quite remarkably that the gap
size distribution was well represented by the spacing distribution
\eqref{E:23}, \eqref{E:24} of GUE. It is an interesting challenge to
develop a microscopic model for the parking problem in \cite{Abu}, analogous to
the model for the bus problem in \cite{BaiBorDeiSui}.

\bigskip

\noindent \textbf{Problem 5.} (Random turns vicious walker model)

In \cite{BaiRai2} the authors proved that, as $N\to\infty$, $d_N$,
the distance traveled by the walker starting from 0, behaves
statistically like the largest eigenvalue of a GOE matrix. More
precisely, they showed that
\begin{equation}\label{E:55}
\lim_{N\to\infty} \text{Prob}
\Bigl(\frac{d_N-2\sqrt{N}}{N^{^{1/6}}}\leq t\Bigr)=F_1(t)
\end{equation}
where $F_1$ is given by \eqref{E:32}. In a variant of this model,
\cite{For2}, the walkers again start at 0,1,2,..., and move to the
left for a time $N$; thereafter they must move to the right,
returning to their initial positions 0,1,2,... at time $2N$. Let
$d_N'$ denote the maximum excursion of the walker starting from 0.
Then Forrester shows that $d_N'$ behaves statistically  like the
largest eigenvalue of a GUE matrix,
\begin{equation}\label{E:56}
\lim_{N\to\infty}
\text{Prob}\Bigl(\frac{d_N'-2\sqrt{N}}{N^{^{1/6}}}\leq
t\Bigr)=F_2(t)
\end{equation}
where $F_2$ is given by \eqref{E:28}.

The proofs of \eqref{E:55} and \eqref{E:56} rely on the observation
of Forrester in \cite{For2} that, in the first case, the set of walks
is in one-to-one correspondence with the set $Y^{(1)}$ of standard
Young tableaux of size $N$ (see \cite{Sag}), whereas in the second
case, the variant model, the set of walks is in one-to-one
correspondence with the set $Y^{(2)}$ of pairs $(P,Q)$ of standard
Young tableaux of size $N$ with the same shape,
$\text{sh}(P)=\text{sh}(Q)$. In both cases, $d_N$ and $d_N'$ equal the
number of boxes in the first row of the corresponding standard Young tableaux.
Uniform measure on $Y^{(2)}$ (resp $Y^{(1)}$) gives rise to
Plancherel measure (resp. $\text{Prob}^\text{(inv)}$) on Young
diagrams of size $N$, and the proof of \eqref{E:56} then follows
from \eqref{E:41}, \eqref{E:45}, and the proof of \eqref{E:55}
follows from \eqref{E:46}.

In \cite{Bai}, Baik proved the analogue of \eqref{E:55},
\eqref{E:56} for the so-called lock step vicious walker introduced in
\cite{Fis}. The proof in \cite{Bai} relies in part on an observation
of Guttmann et al. in \cite{GutOwcVie}, which preceded \cite{For2},
that the set of path configurations for the lock step model is in
one-to-one correspondence with the set of semi-standard Young
tableaux  (see \cite{Sag}).

\bigskip

\noindent \textbf{Problem 6.} (Aztec diamond)

After scaling by $n+1$, Jockush et al., \cite{JocProSho}, considered the tiling problem with
dominos of size $\frac{2}{n+1}\times\frac{1}{n+1}$ in the tilted square $\mathbf{T}_0=\{(u,v)\,:\,|u|+|v|\leq1\}$.
As $n\to\infty$, they found that the inscribed circle $C_0=\{(u,v)\,:\,u^2+v^2=\frac12\}$, which they called
the \textit{arctic circle}, plays a remarkable role. In the four regions of $\mathbf T_0$ outside $C_0$, which they call
the \textit{polar regions} and label N, E, S, W clockwise from the top, the typical tiling is
\textit{frozen}, with all the dominoes in N and S horizontal, and all the dominos in E and W vertical.
In the region inside $C_0$, which they call the \textit{temperate zone}, the tiling is random.
(See, for example, \texttt{http://www.math.wisc.edu/$\sim$propp/tiling}, where a tiling with $n=50$ is displayed.)

But more is true. In \cite{Joh1}, \cite{Joh2}, Johansson considered  fluctuations of the boundary of the
temperate zone about the circle $C_0$. More precisely, for $-1<\alpha<1$, $\alpha\neq0$, let
$$
(x_\alpha^+,y_\alpha^+)=\bigl(\tfrac{\alpha+\sqrt{1-\alpha^2}}{2},\tfrac{\alpha-\sqrt{1-\alpha^2}}{2}\bigr),
\quad
(x_\alpha^-,y_\alpha^-)=\bigl(\tfrac{\alpha-\sqrt{1-\alpha^2}}{2},\tfrac{\alpha+\sqrt{1-\alpha^2}}{2}\bigr)
$$
denote the two points of intersection of the line $u+v=\alpha$ with $C_0=\{u^2+v^2=\frac12\}$. Then for fixed $\alpha$,
Johansson showed that the fluctuations of the boundary of the temperate zone along the line $u+v=\alpha$ about the points
$(x_\alpha^+,y_\alpha^+)$ and $(x_\alpha^-,y_\alpha^-)$
were described by the Tracy-Widom distribution $F_2$ (see \cite{Joh2}, equation~(2.72), for a precise statement).
Johansson proceeds by expressing the fluctuations in terms of the Krawtchouk ensemble (cf. Problem~4), which he then
evaluates asymptotically as $n\to\infty$. Such an analysis is possible because the associated Krawtchouk polynomials
have an integral representation which can be evaluated asymptotically using the classical method of steepest descent.
In \cite{CohLarPro}, the authors considered tilings of hexagons of size $n$ by unit rhombi and proved an arctic circle
theorem for the tilings as $n\to\infty$ as in the case of the Aztec diamond. In \cite{Joh1}, \cite{Joh2}, Johansson
again expressed the fluctuations of the arctic circle for the hexagons in terms of a random particle ensemble, but now using the Hahn polynomials
rather than the Krawtchouk polynomials. The Hahn polynomials, however, do not have a convenient integral
representation and their asymptotics cannot be evaluated by classical means. This obstacle was overcome by
Baik et al., \cite{BaiKriMcLMill}, who extended the Riemann-Hilbert/steepest descent method in \cite{DKMVZ1}
and \cite{DKMVZ2} to a general class of discrete orthogonal polynomials. In this way they were able to compute
the asymptotics of the Hahn polynomials and verify $F_2$-behavior for the fluctuations of the temperate zone, as in the case of
the Aztec diamond.

\bigskip

\noindent \textbf{Problem 7.} (Airline boarding)

\begin{figure}[h]
\begin{center}
\epsfig{file=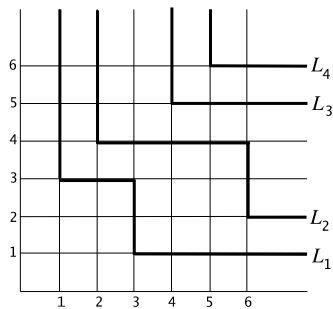,width=1.7in}
\end{center}
\caption{Shadow lines for $\pi\,:\,3\,4\,1\,5\,6\,2$ in $S_6$.}\label{F:airline}
\end{figure}

In \cite{BacBerSapSkiSto} the authors show that $b_N(\pi)$, the boarding time for $N$ passengers
subject to the protocol (a)(b)(c) in Problem~7, behaves statistically like the largest eigenvalue
of a GUE matrix,
\begin{equation}\label{E:57}
\lim_{N\to\infty} \text{Prob}\Bigl(\frac{b_N-2\sqrt{N}}{N^{^{1/6}}}\leq t\Bigr)=F_2(t).
\end{equation}
The proof of \eqref{E:57} in \cite{BacBerSapSkiSto} relies on the description of the Robinson-Schensted
correspondence in terms of Viennot diagrams (see \cite{Sag}). We illustrate the situation with the permutation
$\pi\,:\,3\,4\,1\,5\,6\,2$ in $S_6$ (cf. Problem~3 and \eqref{E:37}).
We say that a point $(x',y')$ \textit{lies
in the shadow} of a point $(x,y)$ in the plane if $x'>x$ and $y'>y$. Plot $\pi$
as a graph $(1,3),(2,4),\ldots,(6,2)$ in the first quadrant of $\IR^2$. Consider all the points
in the graph which are not in the shadow of any other point: in our case $(1,3)$ and $(3,1)$.
The \textit{first shadow line} $L_1$ is the boundary of the combined shadows of these two points (see Figure~\ref{F:airline}).
To form the \textit{second shadow line} $L_2$, one removes the points $(1,3)$, $(3,1)$ on $L_1$, and repeats the
procedure, etc. Eventually one obtains $k_N(\pi)=k$ shadow lines $L_1, L_2, \ldots, L_k$ for
some integer $k$. In our example $k=4$, which we note is precisely $l_6(\pi)$, the length of the longest
increasing subsequence for $\pi$. This is no accident: for any $\pi\in S_N$, we always have $k_N(\pi)=l_N(\pi)$ (see \cite{Sag}).

The beautiful fact is that each shadow line describes a step in the boarding process. Indeed, noting the $y$-values
of the $L_j$'s, we observe that
\begin{equation*}
\begin{split}
L_1\quad &\longleftrightarrow\quad \text{3 and 1 are seated}\\
L_2\quad &\longleftrightarrow\quad \text{4 and 2 are seated}\\
L_3\quad &\longleftrightarrow\quad \text{5 is seated}\\
L_4\quad &\longleftrightarrow\quad \text{6 is seated}\\
\end{split}
\end{equation*}
Thus $b_N(\pi)=k_N(\pi)=l_N(\pi)=q_N(\pi)$, and \eqref{E:57} follows from \eqref{E:41}. In the language of
physics, if we rotate the Viennot diagram for $\pi$ counterclockwise by $45^\text{o}$, we see that the shadow
region of a point on the graph is simply the forward light cone based at that point (speed of light=1). In other
words, for appropriate coordinates $a,b$ we are dealing with the Lorentzian metric $ds^2=dadb$. In order to incorporate
more realistic features into their boarding model, such as the number of seats per row, average amount
of aisle length occupied by a passenger, etc., the authors in \cite{BacBerSapSkiSto} observe that it is enough simply to
replace $ds^2=dadb$ by a more general Lorentzian metric $ds^2=4D^2p(a,b)(dadb+k\alpha(a,b)da^2)$ for
appropriate parameters/functions $D,p,k$ and $\alpha$ (see \cite{BacBerSapSkiSto}, equation~(1)). Thus the basic phenomenon
of blocking in the airline boarding problem is modeled in the general case by the forward light cone
of some Lorentzian metric.

\bigskip

Problems 1 and 2 above, as opposed to 3--7, are purely deterministic and yet it seems
that they are well described by a random model, RMT. At first blush, this might seem counterintuitive,
but there is a long history of the description of deterministic systems by random models.
After all, the throw of a (fair) 6-sided die through the air is completely described by Newton's laws:
Nevertheless, there is no doubt that the \textit{right} way to describe the outcome is probabilistic, with a
one in six chance for each side. With this example in mind, we may say that Wigner was looking for the right ``die''
to describe neutron scattering.

Problems 1--7 above are just a few of the many examples now known of mathematical/physical systems which exhibit
random matrix type universal behavior. Other systems, from many different areas, can be found for example in
\cite{Meh} and the reviews \cite{TW4}, \cite{For3}, and \cite{FerPra}.
A particularly fruitful development has been the discovery of connections between random matrix
theory and stochastic growth models in the KPZ class (\cite{PraSpo}, \cite{FerPra}), and between random
matrix theory and equilibrium crystals with short range interactions (\cite{CerKen}, \cite{FerSpo}, \cite{OkoRes},
\cite{FerPraSpo}).
In addition, for applications to principal component analysis
in statistics in situations where the number of variables is comparable to the sample size, see
\cite{John} and \cite{BBP} and the references therein. For a relatively recent review of
the extensive application of RMT to quantum transport, see \cite{Bee}.

Returning to Wigner's introduction of random matrix theory into theoretical physics,
we note that GOE is of course a mathematical model far removed from the laboratory of
neutrons colliding with nuclei. Nevertheless, Wigner posited that these two worlds were related: With hindsight, we may see
Wigner's insight as a prophetic vision of a scientific commonality across the far borders of physics and mathematics.

\section{Section 5}\label{S:5}

As is clear from the text, many different kinds of mathematics are needed to analyze
Problems~3--7. These include
\begin{itemize}
 \item combinatorial identities
 \item Riemann-Hilbert methods
 \item Painlev\'e theory
 \item theory of Riemann surfaces
 \item representation theory
 \item classical and Riemann-Hilbert steepest descent methods
\end{itemize}
and, most importantly,
\begin{itemize}
\item random matrix theory.
\end{itemize}
The relevant combinatorial identities are often obtained by analyzing random particle systems conditioned not
to intersect, as in Problem~4. The Riemann-Hilbert steepest descent method has its origins in the theory
of integrable systems, as in \cite{DeiZho1}. There is no space in this article to describe the
implementation of any of the above techniques in any detail. Instead, we refer the reader to \cite{Dei2},
which is addressed to a general mathematical audience, for a description of the proof of \eqref{E:41} in particular, using Gessel's
formula in combinatorics \cite{Ges}, together with the Riemann-Hilbert steepest descent method.
For Problem~2 the proofs are based on combinatorial facts and random matrix theory, together with
techniques from the theory of L-functions, over $\IQ$ and also (in \cite{KatSar}) over finite fields.

Universality as described in this article poses a challenge to probability theory per se. The central limit theorem \eqref{E:2}
above has three components: a statistical component (take independent, identically distributed random
variables, centered and scaled), an algebraic component (add the variables), and an analytic component (take the limit in
distribution as $n\to\infty$). The outcome of this procedure is then universal - the Gaussian distribution.
The challenge to probabilists is to describe an analogous purely probabilistic procedure whose outcome is
$F_1$, or $F_2$, etc. The main difficulty is to identify the algebraic component, call it operation $X$. Given $X$,
if one takes i.i.d.'s, suitably centered and scaled, performs operation $X$ on them,
and then takes the limit in distribution, the outcome should be $F_1$, or $F_2$, etc. Interesting progress has
been made recently (see \cite{BodMar} and \cite{BaiSui}) on identifying $X$ for $F_2$. For a different approach to
the results in \cite{BodMar} and \cite{BaiSui}, see \cite{Sui}, where the author uses a very interesting
generalized version of the Lindeberg principle due to Chatterjee~\cite{Cha1}, \cite{Cha2}.

Our final comment/speculation is on the space $D$, say, of probability distributions. A priori, $D$ is just a set
without any ``topography''. But we know at least one interesting point on $D$, the Gaussian distribution $F_{_G}$.
By the central limit theorem, it lies in a ``valley'', and nearby distributions are drawn towards it.
What we seem to be learning is that there are other interesting distributions, like $F_1$ or $F_2$, etc.,
which also lie in ``valleys'' and draw nearby distributions in towards them. This suggests that we equip $D$
with some natural topological and Riemannian structure, and study the properties of $D$ as a manifold per se.


\end{document}